\documentclass[aps,prl,preprint,amsmath,amssymb,showpacs,superscriptaddress]{revtex4-2}

\usepackage{graphicx}% Include figure files
\usepackage{dcolumn}% Align table columns on decimal point
\usepackage{bm}
\usepackage{graphicx}
\usepackage{overpic}
\usepackage{longtable}
\usepackage{epsfig}
\usepackage{epsf}
\usepackage{booktabs}
\usepackage{colortbl}
\usepackage{booktabs}
\usepackage{amssymb}
\usepackage{amsmath}
\usepackage{amsthm}
\usepackage{mathtools}
\usepackage{braket}
\usepackage{multirow}
\usepackage{verbatim}
\usepackage[T1]{fontenc}
\usepackage{amsfonts}
\usepackage{scalerel}
\usepackage{cases}
\usepackage{pgfplots}
\usepackage{float}
\usepackage{lipsum}
\usepackage{gensymb}
\usepackage[colorlinks=true,linkcolor=blue,citecolor=blue]{hyperref}
\usepackage{mathrsfs}
\usepackage{siunitx}
\usepackage{lineno}
\newcommand{\myfont}{\fontfamily{phv}\selectfont}
\newcommand{\figfont}{\myfont \textsf \large}

\makeatletter
\begin{document}

\title{Observation of measurement-induced localization phenomena in a solid-state spin system}

\author{Haolin Zhang}
\thanks{These authors contributed equally to this work.}
\affiliation{Laboratory of Spin Magnetic Resonance, School of Physical Sciences, Anhui Province Key Laboratory of Scientific Instrument Development and Application, University of Science and Technology of China, Hefei 230026, China}
\affiliation{Hefei National Research Center for Physical Sciences at the Microscale, University of Science and Technology of China, Hefei 230026, China}
\affiliation{Hefei National Laboratory, University of Science and Technology of China, Hefei 230088, China}

\author{Shaoyi Xu}
\thanks{These authors contributed equally to this work.}
\affiliation{Laboratory of Spin Magnetic Resonance, School of Physical Sciences, Anhui Province Key Laboratory of Scientific Instrument Development and Application, University of Science and Technology of China, Hefei 230026, China}

\author{Zihang Zheng}
\thanks{These authors contributed equally to this work.}
\affiliation{Laboratory of Spin Magnetic Resonance, School of Physical Sciences, Anhui Province Key Laboratory of Scientific Instrument Development and Application, University of Science and Technology of China, Hefei 230026, China}
\affiliation{Hefei National Laboratory, University of Science and Technology of China, Hefei 230088, China}

\author{Mengqi Wang}
\affiliation{Laboratory of Spin Magnetic Resonance, School of Physical Sciences, Anhui Province Key Laboratory of Scientific Instrument Development and Application, University of Science and Technology of China, Hefei 230026, China}
\affiliation{Hefei National Research Center for Physical Sciences at the Microscale, University of Science and Technology of China, Hefei 230026, China}

\author{Mingzhe Liu}
\affiliation{Laboratory of Spin Magnetic Resonance, School of Physical Sciences, Anhui Province Key Laboratory of Scientific Instrument Development and Application, University of Science and Technology of China, Hefei 230026, China}

\author{Zhiyuan Zhao}
\affiliation{Laboratory of Spin Magnetic Resonance, School of Physical Sciences, Anhui Province Key Laboratory of Scientific Instrument Development and Application, University of Science and Technology of China, Hefei 230026, China}
\affiliation{Hefei National Research Center for Physical Sciences at the Microscale, University of Science and Technology of China, Hefei 230026, China}

\author{Ya Wang}
\affiliation{Laboratory of Spin Magnetic Resonance, School of Physical Sciences, Anhui Province Key Laboratory of Scientific Instrument Development and Application, University of Science and Technology of China, Hefei 230026, China}
\affiliation{Hefei National Research Center for Physical Sciences at the Microscale, University of Science and Technology of China, Hefei 230026, China}
\affiliation{Hefei National Laboratory, University of Science and Technology of China, Hefei 230088, China}

\author{Tianyu Xie}
\email{xie1021@ustc.edu.cn}
\affiliation{Laboratory of Spin Magnetic Resonance, School of Physical Sciences, Anhui Province Key Laboratory of Scientific Instrument Development and Application, University of Science and Technology of China, Hefei 230026, China}
\affiliation{Hefei National Research Center for Physical Sciences at the Microscale, University of Science and Technology of China, Hefei 230026, China}

\author{Fazhan Shi}
\email{fzshi@ustc.edu.cn}
\affiliation{Laboratory of Spin Magnetic Resonance, School of Physical Sciences, Anhui Province Key Laboratory of Scientific Instrument Development and Application, University of Science and Technology of China, Hefei 230026, China}
\affiliation{Hefei National Research Center for Physical Sciences at the Microscale, University of Science and Technology of China, Hefei 230026, China}
\affiliation{Hefei National Laboratory, University of Science and Technology of China, Hefei 230088, China}
\affiliation{School of Biomedical Engineering and Suzhou Institute for Advanced Research, University of Science and Technology of China, Suzhou 215123, China}
\affiliation{The First Affiliated Hospital of USTC, Division of Life Sciences and Medicine, University of Science and Technology of China, Hefei 230026, China}

\begin{abstract}
Quantum measurement can profoundly impact the dynamics of many-body systems, for instance, inducing phase transitions from volume-law chaotic phases to area-law localized phases. While chaotic phases emerge naturally in generic many-body systems, the experimental realization of a fully localized phase driven by measurement remains challenging, primarily owing to the stringent requirement of near-perfect isolation of the system from environmental dissipation. Here we experimentally observe measurement-induced localization phenomena in an interacting $^{13}\text{C}$ nuclear spin system in diamond. In contrast to previous circuit-model-based experiments, our well-isolated system evolves under a physical many-body Hamiltonian with measurements intervening at intervals. By implementing either projective measurements on a single spin or global weak measurements on the entire system, we observe substantially prolonged spin autocorrelation times as the measurement frequency increases. Our results, combined with numerical simulations, provide evidence that quantum measurement can drive a many-body system into a localized regime, and lay the groundwork for future experimental investigations into measurement-induced many-body dynamics and related entanglement phenomena.
\end{abstract}

\maketitle

% \section*{Introduction}
The evolution of a quantum system generally scrambles the initial information contained in its non-equilibrium states, a process usually termed thermalization. 
However, this generic tendency can be suppressed when the system is subjected to frequent measurements, which can freeze the evolution through the quantum Zeno effect (QZE)~\cite{Misra1977, Itano1990}.
While previous QZE studies~\cite{Streed2006, Patil2015, Kalb2016, Harrington2017, Pfender2019, Dasari2022} have been mostly focused on single-particle systems, recent works on measurement-induced phase transitions (MIPT) have extended this paradigm to the many-body realm~\cite{Li2018, Skinner2019, Choi2020a, Tang2020, Turkeshi2020, Lang2020, Goto2020, Fuji2020, Alberton2021, Minato2022, Mueller2022, Noel2022, Koh2023, Szyniszewski2023, Hoke2023, Fisher2023, Kamakari2025, Li2025}. 
A central prediction of MIPT is that, with an increasing rate of measurements, the competition between projective measurements and unitary evolution can lead to a transition from a volume-law entangled phase to a disentangled “quantum Zeno phase” where the entanglement entropy saturates in an area-law~\cite{Li2018}.
Subsequently, more unconventional behaviors have been identified with logarithmically and algebraically scaling entanglement entropy~\cite{Skinner2019, Tang2020, Alberton2021, Mueller2022}.

Experimental exploration of MIPT requires near-perfect isolation of the experimental platform from the environment.
The isolation ensures that a quantum trajectory remains well-defined and that the hallmarks of localized phases can be observed unambiguously. 
Many experimental systems have been suggested for studying MIPT~\cite{Fisher2023, Li2025}, including superconducting circuits~\cite{Koh2023, Hoke2023, Kamakari2025}, trapped ions~\cite{Noel2022} and cold atoms~\cite{Goto2020}, yet they all face the isolation challenge.
Superconducting circuits are vulnerable to electromagnetic noise, while trapped ions and cold atoms contend with issues such as heating and state leakage.
Among solid-state alternatives, ensemble nitrogen-vacancy (NV) centers in diamond have been widely used to study many-body physics~\cite{Choi2017, Choi2017a, Kucsko2018, Zu2021, Davis2023, Martin2023}, but they suffer from relatively short coherence times and low readout fidelity.

In this work, instead of relying on random gate sequences as in previous MIPT experiments~\cite{Noel2022, Koh2023, Kamakari2025}, we investigate a solid-state many-body system governed by its intrinsic dipolar Hamiltonian: the $^{13}\text{C}$ nuclear spin network surrounding an NV center in diamond.
Shielded by the rigid diamond lattice, these $^{13}\text{C}$ spins form stable qubits, and meanwhile they are well isolated from other lattice degrees of freedom, such as band electrons and lattice vibrations, leading to near-perfect isolation of this system. Furthermore, they can be efficiently initialized, addressed, and projectively read out via the NV electron spin~\cite{Broadway2018, Zhang2019, Neumann2010, Taminiau2012, Abobeih2019, Xie2021a, Xie2023}. 
Experimentally, two types of measurements are employed: projective measurements on the target spin and global weak measurements on the entire network. 
Our results demonstrate that, despite their distinct natures, both types of measurements induce a similar prolongation of the spin autocorrelation time $T_1$ — a phenomenon we identify as the measurement-induced localization. 
For the global weak measurements, we further establish a direct connection between the measurement-induced localization and the localized regime of the MIPT in our system by calculating the subsystem mutual information. 
The mutual information reveals a crossover region which coincides with the onset of $T_1$ prolongation. Consequently, a significantly prolonged $T_1$ places our system in a measurement-induced localized regime.

\begin{figure}
    \centering
    \begin{overpic}[width=1.0\linewidth]{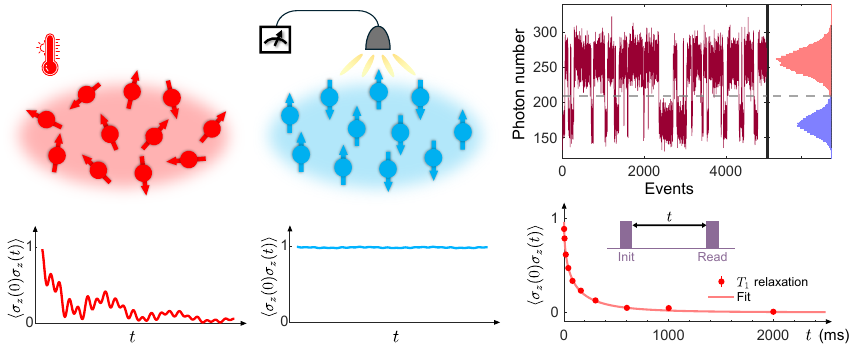}
    \put (0, 41) {\figfont{(a)}}
    \put (30, 41) {\figfont{(b)}}
    \put (59, 41) {\figfont{(c)}}
    \put (59, 17) {\figfont{(d)}}
    \end{overpic}
    \caption{
    Spin localization induced and monitored by external measurements.  (a) Schematic of the $^{13}\text{C}$ spin system in diamond and thermalization.
    When unmonitored, the system thermalizes naturally with a fast-decaying spin autocorrelation function. 
    (b) In the limit of infinitely frequent measurements, the spin autocorrelation function is non-decaying. 
    (c) Projective measurements of the target $^{13}\text{C}$ spin state. 
    Each measurement consists of 1100 repetitions of a $\text{C}_n \text{NOT}_e$ gate followed by a laser pulse.
    The left line graph shows data from 5000 measurements; the histogram on the right corresponds to 55000 measurements.
    The gray dashed line denotes the threshold used to determine the nuclear spin state.
    (d) Unmonitored relaxation of the target spin.
    The process is depicted by a stretched exponential relaxation with a typical time of 77.9(2.6) ms and a stretching exponent of 0.539(16). 
    The inset shows the applied sequence. 
    Error bars represent $\pm 1$ standard deviation (SD).
    }
    \label{fig:1}
\end{figure}

% \section*{System and measurement}
The $^{13}\text{C}$ nuclear spins in diamond constitute an interacting many-body system under magnetic dipolar interactions:
\begin{align}
H_\text{n,int}\approx \sum_{i<j}J_{ij}[\frac{1}{2}(I^i_+I^j_- + I^i_-I^j_+) - 2I^i_z I^j_z],
\end{align}
where the secular approximation has been applied in the high-field limit. 
The dipolar coupling constant $J_{ij}\propto r_{ij}^{-3}$ is a function of the relative geometric positions $r_{ij}$ of the spin pair $(i,j)$;
$I_z$ is the spin $z$ operator, 
and $I_{+(-)}$ is the spin raising (lowering) operator. 
Under this Hamiltonian, the spin system stays in a volume-law chaotic phase and naturally evolves towards its thermal equilibrium (Fig.~\hyperref[fig:1]{\ref*{fig:1}(a)}). 
When subject to frequent measurements, the system becomes localized, preserving local quantum information in the limit of infinitely frequent measurements (Fig.~\hyperref[fig:1]{\ref*{fig:1}(b)}).
To observe this measurement-induced localization (MIL), the spin states are read out at different time intervals to track the spin autocorrelation function $\braket{\sigma_z (0)\sigma_z (t)}$ ($\sigma_z=\ket{\uparrow}\bra{\uparrow}-\ket{\downarrow}\bra{\downarrow}$ is the Pauli-$z$ operator). 
We note that the concept of MIL was originally introduced in the context of an ultracold lattice gas, where measurement suppresses tunneling and localizes the gas in real space~\cite{Patil2015}. Here we use MIL to denote the measurement-induced suppression of thermalization and prolongation of autocorrelation time in a many-body spin system.

The spin-state readout is achieved by employing a technique adapted from previous works~\cite{Neumann2010, Xie2021a}, where the nuclear spin state is repetitively transferred to and read out via an adjacent NV electronic spin.
The NV center, when staying in its negative charge state ($\ket{\mathrm{NV}^-}$), can be efficiently read out and initialized into the $\ket{m_S=0}$ ground-state sublevel via green laser pulses.
By repeatedly applying a sequence consisting of a $\text{C}_n \text{NOT}_e$ gate (an electron NOT gate conditional on the nuclear state) followed by a laser pulse, and collecting the fluorescence photons emitted from the NV center, nuclear spin states can be distinguished (Fig.~\hyperref[fig:1]{\ref*{fig:1}(c)}). 
The key to selectively conducting such $\text{C}_n \text{NOT}_e$ gates is the hyperfine interaction between the target nuclear spin and the NV electronic spin. 
The transverse components of the hyperfine interaction are suppressed by the electron Zeeman term, reducing the Hamiltonian to its secular approximation: $H_{\text{hyp}} \approx \sum_{k=x,y,z} A_{zk}S_z I_k$, where $A_{zk}$ denotes the relevant components of the hyperfine coupling tensor linking the NV spin $z$ operator $S_z$ to the $^{13}\text{C}$ spin operator $I_k$. 

% \section*{Thermalization}
Intrinsic thermalization of nuclear spins establishes a baseline for our study. 
To characterize it, the autocorrelation function introduced above is measured for a single $^{13}\text{C}$ spin: after initialization, its state is read out following a variable free evolution time $t$.
Notably, the projective readout at the end of each experimental cycle simultaneously re-initializes the spin for the subsequent one.
In order to solely study the unmonitored thermalization dynamics, a charge-state readout of the NV center is additionally performed during the initialization step and only the $\ket{\mathrm{NV}^-}$ events are post-selected (see Supplemental Material).
The obtained decay curve (Fig.~\hyperref[fig:1]{\ref*{fig:1}(d)}) is well described by a stretched exponential function, yielding a characteristic decay time of 77.9(2.6) ms and a stretching exponent of 0.539(16). 
The origin of this particular stretching exponent remains an open question, which may be linked to the many-body glassy dynamics previously reported in spin ensembles~\cite{Signoles2021} and deserves further investigations.

\begin{figure}
    \centering
    \begin{overpic}[width=1.0\linewidth]{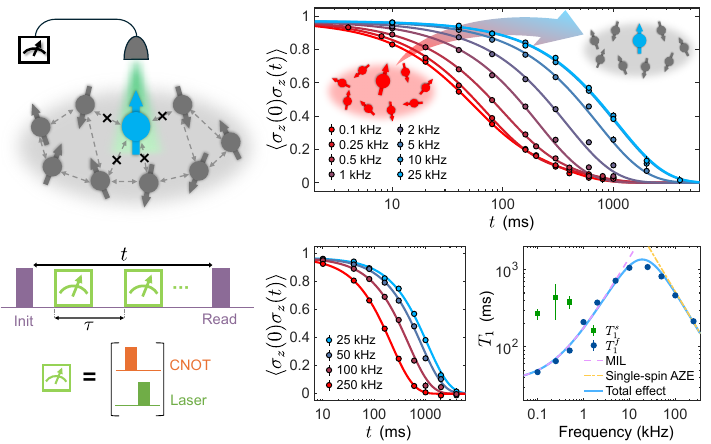}
    \put (0, 61.5) {\figfont{(a)}}
    \put (0, 28) {\figfont{(b)}}
    \put (37.5, 61.5) {\figfont{(c)}}
    \put (37.5, 28) {\figfont{(d)}}
    \put (67, 28) {\figfont{(e)}}
    \end{overpic}
    \caption{
    Projective measurements on one $^{13}\text{C}$ spin. 
    (a) Schematic of the single-spin localization.
    (b) Sequence to induce and observe the localization. 
    (c) Localization phenomena induced by projective measurements at different frequencies.
    (d) The anti-Zeno effect of single spins under frequent measurements. 
    (e) Exponential decay time constants ($T_1^s, T_1^f$) extracted from fitting the spin autocorrelation curves.
    Frequency is defined as the reciprocal of $\tau$.
    The MIL curve indicates the spin localization phenomena induced by measurements.
    All error bars represent $\pm 1$ SD.
    }
    \label{fig:2}
\end{figure}

% \section*{Localization induced by projective measurements}
Building on this baseline, we next examine whether frequent projective measurements can completely suppress the interaction between the target $^{13}\text{C}$ spin and its surrounding $^{13}\text{C}$ spins (Fig.~\hyperref[fig:2]{\ref*{fig:2}(a)}).
In this modified protocol, a series of such measurements are performed on the target spin during its free evolution, prior to the final readout (Fig.~\hyperref[fig:2]{\ref*{fig:2}(b)}).
To accelerate the experiment, each projective measurement is implemented via a single $\text{C}_n \text{NOT}_e$ gate, which entangles the target spin with the NV electronic spin: 
$(\alpha\ket{\uparrow}+\beta\ket{\downarrow}) \otimes \ket{m_S=0} \Rightarrow \alpha\ket{\uparrow} \ket{m_S=0} + \beta\ket{\downarrow} \ket{m_S=+1}$, followed by a laser pulse that re-initializes the NV center to $\ket{m_S=0}$. 
This process destroys the entanglement between the target spin and other spins, thereby constituting an effective projective measurement on the target spin.
From an information-theoretic perspective, the NV center emits enough photons carrying information to projectively determine the $^{13}\text{C}$ spin state, even though not all photons are collected. 
The experimental results confirm that increasing the measurement frequency can better disentangle the target $^{13}\text{C}$ spin from its nuclear spin environment, extending its spin autocorrelation time up to $\SI{1.06}{\second}$ (Fig.~\hyperref[fig:2]{\ref*{fig:2}(c)}).
Notably, the obtained curves at low measurement frequencies are better fitted by a double-exponential function, showing a separation of relaxation timescales — a faster component associated with local equilibration among several strongly coupled neighboring spins, followed by a slower component associated with the rest of the network.
As shown in Fig.~\hyperref[fig:2]{\ref*{fig:2}(d)}, excessively frequent measurements accelerate the relaxation, due to the anti-Zeno effect~\cite{Kofman2000} (AZE) of single spins.
This AZE originates from the transverse hyperfine coupling $A_\perp = \sqrt{A_{zx}^2 + A_{zy}^2}$ between the target spin and the NV center, which introduces off-diagonal terms in the Hamiltonian.
The periodic NV re-initialization thus induces additional single-spin relaxation, independent of many-body spin dynamics.
We note that this unintended AZE can be mitigated by increasing the external magnetic field, which quadratically suppresses the effect of off-diagonal terms.

% \section*{Localization induced by global weak measurements}
Intriguingly, the laser pulse itself constitutes an effective weak measurement of the target $^{13}\text{C}$ spin: the pulse drives the transition among the NV spin sublevels, thus inducing decoherence on the target spin via the hyperfine coupling term $A_{zz}S_z I_z$.
Conceptually, the $^{13}\text{C}$ spin state is encoded in the energy shift of the NV sublevels, and by optically driving these transitions, a laser pulse indirectly probes the $^{13}\text{C}$ spin.
Moreover, for $^{13}\text{C}$ spins with appreciable longitudinal couplings $A_\parallel = A_{zz}$ to the NV spin, a single laser pulse effectively implements a global weak measurement on them with a uniform measurement strength $\gamma_\text{global}\approx 0.1 f$ that can be further tuned by varying the laser pulse repetition frequency $f$ (see Supplemental Material).
In this work, we refer to these spins collectively as the spin network, defined operationally by $A_\parallel/(2\pi) \gtrsim 1\, \text{kHz}$, which includes $\sim$ 100 spins.

\begin{figure}
    \centering
    \begin{overpic}[width=0.65\linewidth]{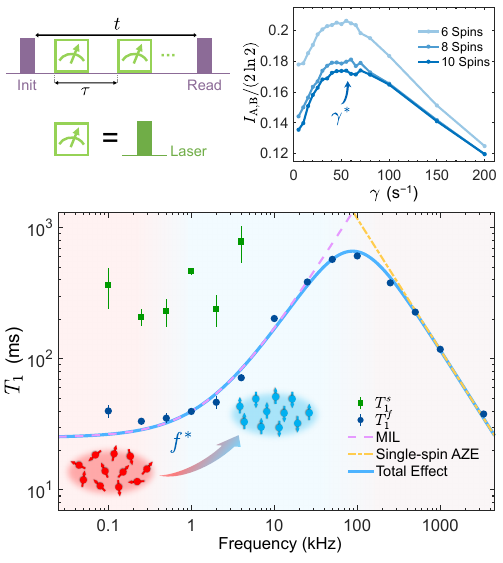}
    \put (0.5, 100) {\figfont{(a)}}
    \put (43, 100) {\figfont{(b)}}
    \put (0.5, 63) {\figfont{(c)}}
    \end{overpic}
    \caption{
    Global weak measurements on the $^{13}\text{C}$ spin network.
    (a) Sequence to perform global weak measurements. 
    Each measurement only contains a 300-\text{ns} laser pulse. 
    (b) Steady-state values of mutual information $I_\text{A,B}$ of two subsystems of the spin network at different measurement strengths. The data show simulation results for three system sizes: $N_\text{spin}=$ 6, 8 and 10.
    (c) Exponential decay time constants ($T_1^s, T_1^f$) extracted from fitting the obtained spin autocorrelation curves (see Supplemental Material). 
    The repetition frequency is defined as the reciprocal of $\tau$. Background shading denotes three regimes: thermalized, localized and AZE-dominated. The blurred bands at the crossovers indicate their uncertainty.
    }
    \label{fig:3}
\end{figure}

The ability to perform global weak measurements motivates a natural extension of our experiment: we replace the sequence of projective measurements with a sequence of single laser pulses (Fig.~\hyperref[fig:3]{\ref*{fig:3}(a)}), which allows us to investigate the emergence of localization induced purely by global weak measurements.
Since this experiment contains all the necessary ingredients for an MIPT, we further explore it by calculating the mutual information between two small subsystems A and B of the spin network in a steady state, defined as $I_\text{A,B}=S_\text{A}+S_\text{B}-S_{\text{A}\cup\text{B}}$, where $S_\text{X}=-\text{Tr}[\rho_\text{X} \ln \rho_\text{X}]$ is the von Neumann entropy of subsystem X.
In this work, A and B are taken to be individual spins.
To define the steady state, the entanglement entropy $S(t)$ is calculated to  identify the time window over which it saturates; all mutual information results are averaged over this window (see Supplemental Material). 
Fig.~\hyperref[fig:3]{\ref*{fig:3}(b)} shows the same non-monotonic behavior of mutual information as previous works~\cite{Li2019, Fuji2020} for a strongly-coupled pair of spins: it grows with measurement strength $\gamma$ in the weakly monitored regime, draw a peak at a critical value $\gamma^*$, and decreases when the measurement dominates.
We therefore interpret the critical $\gamma^*$ in Fig.~\hyperref[fig:3]{\ref*{fig:3}(b)} as an estimate of the crossover, and regard $\gamma>\gamma^*$ as a rough criterion for our spin network to be in the localized regime.

However, the mutual information is not experimentally accessible in the current platform, so we instead measure the autocorrelation function and compare it with our established localized-regime criterion.
Using $\gamma_\text{global}\approx 0.1 f$, we convert the crossover $\gamma^*$
into an equivalent repetition frequency $f^*$.
The results are similar to the projective-measurement case: at low measurement frequencies, faster measurements prolong the spin autocorrelation time; beyond an optimal frequency, the AZE dominates the dynamics and promotes relaxation (Fig.~\hyperref[fig:3]{\ref*{fig:3}(c)}).
The minor decrease of $T_1$ at low frequencies is likely driven by the increased laser-induced polarization of the $\ket{m_S=0}$ state — compared to an unpolarized NV spin state, this polarized state generates a disorder-free Hamiltonian under which the nuclear spins tend to thermalize.
Crucially, the onset of the $T_1$ enhancement falls within the crossover region extracted from the mutual information simulation, which establishes a direct connection between MIL and the MIPT localized regime in our system: a significant enhancement of $T_1$ can serve as an experimental indicator of the localized regime. 
The subsequent decrease of $T_1$ caused by the single-spin AZE is at orders of magnitude higher frequencies and thus does not influence the existence of the localized regime.
We emphasize that the specific $^{13}\text{C}$ spin we read out is not fundamentally different from the others in the network, since it experiences the same global weak measurement and nuclear spin-spin interactions comparable to those of dozens of other spins. 
Consequently, it serves as a probe of the collective response of the entire spin network and reveals the generic localization behavior.

\begin{figure}
    \centering
    \begin{overpic}[width=0.8\linewidth]{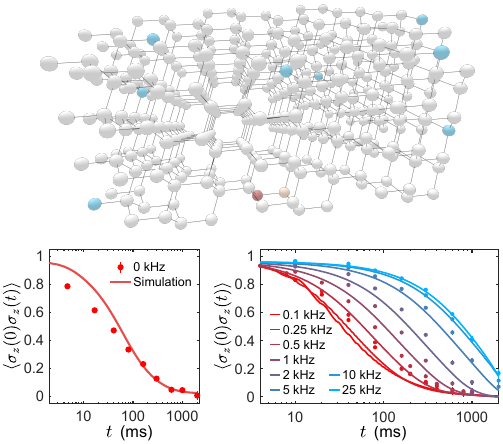}
    \put (0.5, 86) {\figfont{(a)}}
    \put (0.5, 38.5) {\figfont{(b)}}
    \put (42.5, 38.5) {\figfont{(c)}}
    \end{overpic}
    \caption{
    Simulation of the $^{13}\text{C}$ spin dynamics.
    (a) The configuration of eight $^{13}\text{C}$ spins used in simulations, denoted as blue spheres. 
    The red and ivory spheres represent the nitrogen atom and the vacancy in the NV center, respectively.
    (b) Simulation of the relaxation process of the target spin.
    (c) Simulations of spin localization phenomena induced by projective measurements at different frequencies. 
    The data points are experimental results at the corresponding frequencies, shown for comparison.
    All error bars represent $\pm 1$ SD.
    }
    \label{fig:4}
\end{figure}

% \section*{Numerical simulation}
Finally, numerical simulations of the autocorrelation function are performed by solving the Lindblad master equation on a classical computer. 
Since the nuclear spin system is well-isolated, no phenomenological nuclear relaxation or decoherence channels are introduced; therefore, the non-unitarity of the simulated system originates solely from the NV dynamics.
To ensure that the simulations closely reflect the experimental conditions, the coupling parameters ($A_\parallel$, $A_\perp$) of the $^{13}\text{C}$ spins are experimentally obtained by monitoring the NV coherence under dynamical decoupling (DD) sequences~\cite{Taminiau2012, Xie2023}; most NV properties are also determined experimentally. 
The need for a separate time step of the Lindblad equation at each laser pulse, rather than a single unitary evolution, makes the simulation computationally intractable.
Restricted by this complexity, the simulated system includes only the NV center and eight strongly interacting $^{13}\text{C}$ spins (Fig.~\hyperref[fig:4]{\ref*{fig:4}(a)}), whose relative positions with respect to the NV center are determined based on their hyperfine coupling parameters. 
Although eight spins cannot fully capture the many-body dynamics, we select a configuration that most closely matches the experimental data.
Further details are provided in the Supplemental Material.

With these eight spins, the autocorrelation function of the target $^{13}\text{C}$ spin is numerically simulated in the $\ket{\text{NV}^-}$ subspace. 
In the absence of external measurements, the simulation reproduces the observed decay curve (Fig.~\hyperref[fig:4]{\ref*{fig:4}(b)}), reflecting the intrinsic many-body thermalization of the nuclear spin network under dipolar interactions.
When projective measurements are applied to the target spin at varying intervals, the model successfully captures the MIL behavior: the suppression of thermalization leads to an extension of spin autocorrelation time with increasing measurement frequency, in a qualitative agreement with the experimental data (Fig.~\hyperref[fig:4]{\ref*{fig:4}(c)}).
However, the simulations fail to generate the double-exponential behavior in experimental data, possibly owing to the limited spin number.
The simulations of global weak measurements, the anti-Zeno effect, and the $T_1$ shortening due to $\ket{m_S=0}$ state polarization are also shown to agree with experiments and are detailed in the Supplemental Material.
The consistent description of these diverse effects within a single framework provides direct validation of our model.

% \section*{Conclusions}
By observing prolonged spin autocorrelation time $T_1$, we demonstrate measurement-induced localization phenomena in the $^{13}\text{C}$ spin network surrounding an NV center in diamond under both projective and global weak measurements.
Previous MIPT experiments on other platforms all face varying degrees of environmental noise~\cite{Noel2022, Koh2023, Hoke2023, Kamakari2025}.
In contrast, $^{13}\text{C}$ spins in diamond exhibit superior isolation, which is further supported by our numerical simulations that include no additional dissipation and yet reproduce the experimental data.
With such a well-isolated system with intrinsic dipolar interactions, we intend to investigate entanglement structures and transition behaviors, just like circuit-based MIPT experiments~\cite{Koh2023, Hoke2023, Kamakari2025}, but a direct observation of entanglement entropy or mutual information is still beyond the reach of our current platform. We therefore identify a mutual-information crossover and link it to the experimentally accessible $T_1$, so that its observed prolongation can serve as an alternative signature of the MIPT localized regime.
In circuit-based experiments, the postselection overhead~\cite{Fisher2023} has been circumvented by using spacetime duality mapping~\cite{Ippoliti2021, Hoke2023} or cross-entropy benchmarking~\cite{Li2023, Kamakari2025} — the latter of which requires the system to be classically simulable. 
For a many-body system governed by a physical Hamiltonian, such techniques are not directly applicable.

The combination of long relaxation times and a measurement rate swift relative to intrinsic dynamics enables the platform to perform mid-circuit measurements, providing a promising foundation for future studies of measurement-induced entanglement phenomena emerging under genuine many-body Hamiltonians.
Amid the experimental difficulties discussed above, a distinctive advantage of our system is the long-range dipolar interactions among nuclear spins, allowing us to explore novel entanglement phases~\cite{Minato2022, Mueller2022} beyond the reach of short-range models.
At the same time, by preparing the NV spin in the $\ket{m_S=\pm 1}$ sublevels, the hyperfine interaction naturally generates site-dependent disorder for the $^{13}\text{C}$ spins, enabling the exploration of MIPT in a disordered monitored system~\cite{Szyniszewski2023} and bridging the physics of many-body localization and measurement-induced criticality within the same platform. 

\section*{Acknowledgments}
This work was supported by the National Natural Science Foundation of China (grant nos. T2125011, 12274396, and 12404555), the CAS (grant no. YSBR-068), the Quantum Science and Technology-National Science and Technology Major Project (grants nos. 2021ZD0302200 and 2021ZD0303204), the China Postdoctoral Science Foundation (grant nos. 2023M743399 and GZB20240718), and the Fundamental Research Funds for the Central Universities, New Cornerstone Science Foundation through the XPLORER PRIZE.

\newpage
\bibliography{NV_abbr}
\newpage

\end{document}

\title{Supplemental Material: Observation of measurement-induced localization phenomena in a solid-state spin system}

\author{Haolin Zhang}
\thanks{These authors contributed equally to this work.}
\affiliation{Laboratory of Spin Magnetic Resonance, School of Physical Sciences, Anhui Province Key Laboratory of Scientific Instrument Development and Application, University of Science and Technology of China, Hefei 230026, China}
\affiliation{Hefei National Research Center for Physical Sciences at the Microscale, University of Science and Technology of China, Hefei 230026, China}
\affiliation{Hefei National Laboratory, University of Science and Technology of China, Hefei 230088, China}

\author{Shaoyi Xu}
\thanks{These authors contributed equally to this work.}
\affiliation{Laboratory of Spin Magnetic Resonance, School of Physical Sciences, Anhui Province Key Laboratory of Scientific Instrument Development and Application, University of Science and Technology of China, Hefei 230026, China}

\author{Zihang Zheng}
\thanks{These authors contributed equally to this work.}
\affiliation{Laboratory of Spin Magnetic Resonance, School of Physical Sciences, Anhui Province Key Laboratory of Scientific Instrument Development and Application, University of Science and Technology of China, Hefei 230026, China}
\affiliation{Hefei National Laboratory, University of Science and Technology of China, Hefei 230088, China}

\author{Mengqi Wang}
\affiliation{Laboratory of Spin Magnetic Resonance, School of Physical Sciences, Anhui Province Key Laboratory of Scientific Instrument Development and Application, University of Science and Technology of China, Hefei 230026, China}
\affiliation{Hefei National Research Center for Physical Sciences at the Microscale, University of Science and Technology of China, Hefei 230026, China}

\author{Mingzhe Liu}
\affiliation{Laboratory of Spin Magnetic Resonance, School of Physical Sciences, Anhui Province Key Laboratory of Scientific Instrument Development and Application, University of Science and Technology of China, Hefei 230026, China}

\author{Zhiyuan Zhao}
\affiliation{Laboratory of Spin Magnetic Resonance, School of Physical Sciences, Anhui Province Key Laboratory of Scientific Instrument Development and Application, University of Science and Technology of China, Hefei 230026, China}
\affiliation{Hefei National Research Center for Physical Sciences at the Microscale, University of Science and Technology of China, Hefei 230026, China}

\author{Ya Wang}
\affiliation{Laboratory of Spin Magnetic Resonance, School of Physical Sciences, Anhui Province Key Laboratory of Scientific Instrument Development and Application, University of Science and Technology of China, Hefei 230026, China}
\affiliation{Hefei National Research Center for Physical Sciences at the Microscale, University of Science and Technology of China, Hefei 230026, China}
\affiliation{Hefei National Laboratory, University of Science and Technology of China, Hefei 230088, China}

\author{Tianyu Xie}
\email{xie1021@ustc.edu.cn}
\affiliation{Laboratory of Spin Magnetic Resonance, School of Physical Sciences, Anhui Province Key Laboratory of Scientific Instrument Development and Application, University of Science and Technology of China, Hefei 230026, China}
\affiliation{Hefei National Research Center for Physical Sciences at the Microscale, University of Science and Technology of China, Hefei 230026, China}

\author{Fazhan Shi}
\email{fzshi@ustc.edu.cn}
\affiliation{Laboratory of Spin Magnetic Resonance, School of Physical Sciences, Anhui Province Key Laboratory of Scientific Instrument Development and Application, University of Science and Technology of China, Hefei 230026, China}
\affiliation{Hefei National Research Center for Physical Sciences at the Microscale, University of Science and Technology of China, Hefei 230026, China}
\affiliation{Hefei National Laboratory, University of Science and Technology of China, Hefei 230088, China}
\affiliation{School of Biomedical Engineering and Suzhou Institute for Advanced Research, University of Science and Technology of China, Suzhou 215123, China}
\affiliation{The First Affiliated Hospital of USTC, Division of Life Sciences and Medicine, University of Science and Technology of China, Hefei 230026, China}

\maketitle

\tableofcontents

\section*{Diamond sample}

The experiments are performed on a bulk diamond sample containing a natural abundance of $^{13}\text{C}$ ($\approx$ 1.1\,\%), which yields the estimate of $\sim$ 100 spins with $A_\parallel/(2\pi) \gtrsim 1\, \text{kHz}$.
The nitrogen concentration in the diamond is below 5 parts per billion. 
The top surface of diamond is perpendicular to the [100] crystal axis and the lateral surfaces to [110]. 
The diamond is irradiated with 10-MeV electrons to a total dose of approximately $\SI{1.0}{kGy}$. 
The irradiation creates vacancies, which together with the nitrogen impurities form NV centers. 
One of these NV centers is used to control and read out the surrounding $^{13}\text{C}$ spin network. 
To enhance its luminescence rate to approximately $\SI{600}{kcounts/\second}$, a solid immersion lens is fabricated around the NV center.

\section*{Experimental setup}

As illustrated in Fig.~\hyperref[fig:setup]{\ref*{fig:setup}}, the entire experimental apparatus is temporally synchronized by a multichannel pulse controller (Spincore, PBESR-PRO-500), which serves as the central timing unit governing the coordination between the optical and microwave subsystems. 
Within the optical subsystem, an oil objective (Olympus, UPLXAPO100XO, NA 1.45) is employed both to deliver the excitation laser beams and to collect the resulting phonon-sideband fluorescence (650–800 nm) emitted by the NV center. 
A 532-nm laser (green) is utilized for both initialization and readout of the NV spin state, whereas a 594-nm laser (orange) is employed for charge-state readout. 
To suppress laser leakage that could perturb the $\ket{\text{NV}^-}$ charge state or compromise the longitudinal spin relaxation time, the 532-nm beam is directed through an acousto-optic modulator (AOM) (Gooch \& Housego, 97-03046-01) twice prior to entering the objective; similarly, the 594-nm beam passes through another AOM (Gooch \& Housego, 99-48146-11) twice. 
The collected fluorescence is detected by avalanche photodiodes (APDs) (PerkinElmer, SPCM-AQRH-14) and recorded via a data acquisition card (National Instruments, 6612) interfaced with a computer.
The microwave subsystem comprises an arbitrary waveform generator (Keysight, M8195A) and a high-power amplifier (Talent Microwave, TPA-180265-P43-AC, 18–26.5 GHz). 
The generated microwave pulses are delivered through a coplanar waveguide microstructure to coherently manipulate the ground-state triplet of the NV center. 
The entire optical and microwave system is integrated within a temperature-stabilized copper enclosure, housing two three-axis positioning stages: one supports the diamond sample and the objective lens, and the other holds a composite permanent magnet. 
The bias magnetic field provided by the magnet is precisely aligned parallel to the NV symmetry axis by fine-tuning the positioning stage.

\section*{NV charge state readout}

The principle of NV charge state readout is that the 594-nm laser excites $\ket{\text{NV}^-}$, but not $\ket{\text{NV}^0}$.
To single-shot read out the charge state, a 532-nm laser pulse with a duration on the order of microseconds is used to reset the NV, followed by a 4-ms illumination of 594-nm laser, as shown in Fig.~\hyperref[fig:charge]{\ref*{fig:charge}}.
The emitted photons during the 594-nm laser illumination are collected to determine the charge state.

\section*{Data analysis}

As discussed in the main text, at the end of each experiment cycle, the target $^{13}\text{C}$ spin's state is projectively read out, thus being initialized for the next cycle. 
Therefore, it seems that the spin autocorrelation function $C_{zz}(t) = \braket{\sigma_z (0)\sigma_z (t)}$ can be obtained by calculating $\sum_{k=1}^{N_\mathrm{exp}} s_{k}\,s_{k+1}/N_\mathrm{exp}$ of $N_\mathrm{exp}$ experiment cycles, where $s_{k}$ is the spin readout result of the $k$th cycle:
$s_{k}\triangleq+1$ ($-1$) for the $\ket{\uparrow}$ ($\ket{\downarrow}$) state.
However, since the $\text{C}_n \text{NOT}_e$ gate leads to different NV dynamics for the two nuclear spin states, the target spin in the $\ket{\uparrow}$ and $\ket{\downarrow}$ state may possess different correlation times, generating an imbalanced readout counts of $N_\uparrow$ and $N_\downarrow$. 
This imbalance is evident in the raw data corresponding to Fig.\,2(d). 
To circumvent this issue, all the $C_{zz}$ values reported in this work are calculated by
\begin{align}
C_{zz}=\frac{1}{2}\left(\frac{\sum_{k_\uparrow} s_{(k_\uparrow +1)}}{N_\uparrow} - \frac{\sum_{k_\downarrow}s_{(k_\downarrow +1)}}{N_\downarrow} \right),
\end{align}
where $s_{k_{\uparrow\,(\downarrow)}}=+1\,(-1)$ is observed at the end of the $k_{\uparrow\,(\downarrow)}$-th cycle.

The experimentally obtained spin autocorrelation function is fitted using exponential functions.
For Fig.\,1(c), the data is well described by a stretched exponential function, yielding a characteristic relaxation time of
% $T_1 = \mathrm{77.9(2.6)\,ms}$
77.9(2.6)\,ms 
and a stretching exponent of 0.539(16). 
To compensate for the charge-state readout time in the initialization, the experimental data points are shifted to the right by a linear time offset of $\SI{0.69(19)}{ms}$ when plotting, so that the fitted curve starts at a correlation of 0.9646(13), which is the experimentally obtained readout fidelity. 
This offset is substantially shorter than the charge-state readout time of $\SI{4}{ms}$, suggesting that the nuclear spin thermalization is partially suppressed by the readout laser.
For Fig.\,2, data within the measurement frequency range of $[\SI{0.1}{kHz}, \SI{0.5}{kHz}]$ are fitted with a double-exponential function; a single-exponential fit is applied to the remaining data. 
For the global weak measurement experiment (Fig.~\hyperref[fig:laser]{\ref*{fig:laser}}), data within the measurement frequency range of $[\SI{0.1}{kHz}, \SI{4}{kHz}]$ are fitted with a double-exponential function; a single-exponential fit used elsewhere.

\section*{Deduction of the global weak measurement strength}

By illuminating green laser on the NV center, all the $^{13}\text{C}$ spins with an appreciable longitudinal coupling $A_\parallel$ are dephased, therefore a single laser pulse effectively implements a global weak measurement on the entire nuclear spin network.
To determine the measurement strength, a simplified model is examined, in which a $^{13}\text{C}$ spin is coupled to an NV center (in $\ket{\text{NV}^-}$), with the Hamiltonian $H = A_{zz} S_z I_z$. 
Laser pulses arrive as a Poisson process of intensity $\lambda$, and each pulse drives the transition among $\ket{\text{NV}^-}$ sublevels and polarizes the NV to the $\ket{m_S=0}$ state. 
The polarization degree is experimentally determined as $p_{m_S=0} \approx$ 0.9. 
The effect of a single laser pulse is simply described by the transition matrix (in the ordered basis $\{ \ket{m_S =+1}, \ket{m_S =0}, \ket{m_S =-1} \}$)
\begin{align}
T\approx
\begin{pmatrix}
0.1 & 0.9 & 0\\
0.05 & 0.9 & 0.05\\
0 & 0.9 & 0.1
\end{pmatrix}.
\end{align}

Taking into account the stochastic nature of the laser pulses, the transition matrix over an infinitesimal time interval is 
\begin{align}
P(\mathrm{d}t) = (1-\lambda \mathrm{d}t)I + \lambda \mathrm{d}t T = I+Q\mathrm{d}t,
\end{align}
where $Q=\lambda(T-I)$. 

The nuclear spin precesses at different frequencies conditioned on the electron states, i.e.,
\begin{align}
\omega = \left\{
\begin{aligned}
+A_{zz} & \quad(m_S =+1)\\
0 & \quad(m_S =0)\\
-A_{zz} & \quad(m_S =-1)
\end{aligned}
\right..
\end{align}
For a single quantum trajectory, the accumulated nuclear phase is
\begin{align}
\phi(t)=\int_0^t \omega(s) \mathrm{d}s.
\end{align}
Owing to the symmetry of $\{\ket{m_S =+1}, \ket{m_S =-1} \}$, the ensemble-averaged value of $\phi(t) = 0$, so the nuclear-spin coherence is
\begin{align}
C_n(t)=\braket{\ee^{-\ii\phi(t)}}.
\end{align}

For calculation purposes, an auxiliary vector $\vec{v}(t)$ is introduced, whose components are defined by
\begin{align}
v_\alpha(t) = \mathbb{E}[\ee^{-\ii\phi(t)}\mathbf{1}_{\{m_S = \alpha \text{ at time }t\}}]\quad (\alpha=\left\{+1, 0, -1\right\}),
\end{align}
where $\mathbf{1}_{\{...\}}$ is the indicator function, and $\mathbb{E}$ denotes the unconditional expectation of all possible quantum trajectories of the electron.
Under this definition, $C_n(t) = \sum_\alpha {v_\alpha (t)}$. 

After an infinitesimal time interval $\mathrm{d}t$, two processes may occur:

1. The electron may jump according to the transition matrix $Q$, transferring amplitude from one component of $\vec{v}(t)$ to another;

2. While the electron occupies a definite state $\ket{m_S=\alpha}$, the phase $\phi(t)$ advances at the rate $\omega_\alpha$, multiplying $v_\alpha(t)$ by the infinitesimal phase factor $\ee^{-\ii\omega_\alpha\mathrm{d}t}=1-\ii\omega_\alpha\mathrm{d}t + o(\mathrm{d}t)$.
Therefore, 
\begin{align}
\frac{\mathrm{d}\vec{v}(t)}{\mathrm{d}t} = M\vec{v}(t),\quad M=Q-\ii \ \text{diag}(A_{zz}, 0, -A_{zz}),
\end{align}
and
\begin{align}
\vec{v}(t) = \ee^{Mt}\vec{v}(0).
\end{align}

For $A_{zz} \gg \lambda$ (which is true for dozens of nuclear spins), among the three eigenvalues of $M$, two eigenvalues ($\approx -p_{m_S=0}\lambda \pm \ii A_{zz}$) correspond to fast oscillation and decoherence; the slow-decaying one ($\approx -(1-p_{m_S=0})\lambda$) determines the final decoherence envelope of the nuclear spin. Consequently, the effective measurement strength is
\begin{align}
\gamma_\text{global}\approx (1-p_{m_S=0})\lambda \approx 0.1\lambda.
\end{align}
This result agrees with our instinct: if the NV center is fully polarized to $\ket{m_S =0}$, then no electron jump can happen among the $\ket{\text{NV}^-}$ sublevels, and thus the nuclear spins experience no dephasing.

In experiments, periodic laser pulses are applied instead of Poissonian ones for experimental convenience.
It is assumed that the results are similar to those obtained with Poissonian pulses, as several random experimental factors are expected to contribute to the stochasticity of measurements, such as the influence of electron $T_1$, the instability of the pulse controller's clock, etc.

\section*{Resolving $^{13}\text{C}$ spins}

\subsection{DD sequence}

The hyperfine parameters of the $^{13}\text{C}$ nuclear spins can be precisely measured by adopting methods in previous works~\cite{Taminiau2012, Xie2023}.
In the beginning of a quantum sensing sequence, the NV spin is prepared in the superposition state $(\ket{m_S=0}+\ket{m_S=+1})/\sqrt2$ by the initial $\pi/2$ pulse.
The nuclear spins in two subspaces ($\ket{m_S=0}$ and $\ket{m_S=+1}$) then go through different evolution path, denoted by the unitary evolution operator $U_0$ and $U_1$:
\begin{align}
U = \ket{m_S=0}\bra{m_S=0}\otimes U_0 + \ket{m_S=+1}\bra{m_S=+1}\otimes U_1.
\end{align}
After the evolution, the final $\pi/2$ pulse reads out the real part of NV coherence 
\begin{align}
C=\text{Re}[\text{Tr}(U_0\rho_n U_1^\dagger)]\,
\end{align}
where $\rho_n$ is the initial density matrix of the nuclear spins.
The nuclear spins are assumed to be in the maximally mixed state $\rho_n=\otimes_{i=1}^N \mathbb{I}/2$, therefore,
\begin{align}
C=\text{Re}[\text{Tr}(U_0 U_1^\dagger)].
\end{align}
First the DD sequence is employed, where the microwave pulses controlling the NV center are played between the NV state $\ket{m_S=0}$ and $\ket{m_S=+1}$.
For a DD sequence consisting of an even number $N_\pi$ of $\pi$ pulses, $U_0$ and $U_1$ are:
\begin{align}
U_0 = (V_0 V_1^2 V_0)^{N_\pi/2}, U_1 = (V_1 V_0^2 V_1)^{N_\pi/2},
\end{align}
where $V_0$ and $V_1$ are the unit evolution operator of nuclear spins in the subspace $\ket{m_S=0}$ and $\ket{m_S=+1}$, respectively:
\begin{align}
V_0=\text{e}^{-\ii H_0 \tau/2}, 
V_1=\text{e}^{-\ii H_1 \tau/2}.
\end{align}
Next, the Hamiltonian of $N$ nuclear spins is examined. 
For a short experimental time $\tau$, the interaction among the nuclear spins are neglected.
Since the high magnetic field suppresses all terms that are non-commutable to $S_z$, the subspace Hamiltonian is given by
\begin{align}
H_0=\sum_{i=1}^N \omega_L I_z^{i},\; 
H_1=\sum_{i=1}^N (\omega_L + A_{zz}^i) I_z^{i} + A_{zx}^i I_x^{i} + A_{zy}^i I_y^{i},
\end{align}
where $\omega_L$ ($ <0 $ by definition) is the Larmor frequency of the nuclear spin.
The azimuth angles are redundant for calculating the coherence, thus this expression can be further simplified to
\begin{align}
H_0=\sum_{i=1}^N \omega_L I_z^{i},\; 
H_1=\sum_{i=1}^N (\omega_L + A_\parallel^i) I_z^{i} + A_\perp^i I_x^{i},
\end{align}
where $A_\parallel=A_{zz}$, $A_\perp=\sqrt{A_{zx}^2 + A_{zy}^2}$;
or equivalently, 
\begin{align}
H_0=\sum_{i=1}^N \omega_0 I_z^{i},\; 
H_1=\sum_{i=1}^N \omega_1^i \mathbf{n}^i \cdot \mathbf{I}^i,
\end{align}
where $\omega_0 = \omega_L$, 
$\omega_1^i=-\sqrt{(\omega_L+A_\parallel^i)^2+(A_\perp^i)^2}$, 
and the unit vector $\mathbf{n}^i = (A_\perp, 0, \omega_L+A_\parallel^i) / \omega_1^i$.
Since the Hamiltonian and the initial state of nuclear spins are both separable, the NV coherence is simply the multiplication of the contribution from individual nuclear spins, as elaborated in the previous work~\cite{Taminiau2012}:
\begin{align}
C=C_b\prod_{i=1}^N C_i,
\end{align}
where $C_i$ is the contribution from the $i$th nuclear spin. For an even number of $\pi$ pulses,
\begin{align}
C_i= 1-2(n_x^i)^2 \sin^2(\frac{\omega_0 \tau}{4}) \sin^2(\frac{\omega_1^i \tau}{4}) \frac{\sin^2(N_\pi \phi_i/2)}{\cos^2(\phi_i/2)},
\end{align}
where
\begin{align}
\phi_i = \arccos{\left(\cos(\omega_0 \tau/2)\cos(\omega_1^i \tau/2) - n_z^i \sin(\omega_0 \tau/2)\sin(\omega_1^i \tau/2)\right)}.
\end{align}
$C_b$ is the contribution from the residual spin bath, which is regarded as a classical magnetic noise:
\begin{align}
C_b = \exp{\left(-\frac{1}{2}(\frac{\sigma_b N_\pi \tau}{\pi})^2 
\sum_{k=1}^{+\infty}\frac{1}{(2k-1)^2} \text{sinc}^2\left[\frac{N_\pi(\lvert \omega_L \rvert \tau-(2k-1)\pi)}{2}\right]\right)}.
\end{align}
These expressions are used to fit the DD experimental data for resolving the twelve $^{13}\text{C}$ spins and extracting their hyperfine parameters, as depicted in Fig.~\hyperref[fig:DD]{\ref*{fig:DD}}. Notably, the target $^{13}\text{C}$ spin in this work refers to the first spin in Fig.~\hyperref[fig:DD]{\ref*{fig:DD}(d)}.

\subsection{FID sequence}

In the free induction decay (FID) experiment, the components of $U_0$ and $U_1$ are:
\begin{align}
U_0^i=\ee^{-\ii\omega_0 t I_z^i}, 
U_1^i=\ee^{-\ii\omega_1^i t (\mathbf{n}^i \cdot \mathbf{I}^i)}.
\end{align}
To better fit the experimental data, a slight spin polarization $p_{\text{C}}$ of the first $^{13}\text{C}$ spin is introduced, which is induced via the nuclear spin's interaction with the NV center. 
Therefore, the $N$ nuclear spins are not in maximally mixed state, but
\begin{align}
\rho_n=(\frac{1-p_{\text{C}}}{2}\mathbb{I} + p_{\text{C}}\sigma_z)\otimes_{i=2}^N(\mathbb{I}/2).
\end{align}
Calculating the NV coherence $C=\text{Re}[\text{Tr}(U_0\rho_n U_1^\dagger)]$ yields
\begin{align}
C&=C_b\prod_{i=1}^N C_i, \\
C_i &= 
\begin{cases}
(1-p_{\text{C}})a_i - \ii p_{\text{C}} b_i, & i = 1, \\
a_i, & i \neq 1,
\end{cases}
\end{align}

where
\begin{align}
a_i = & \cos(\omega_0 t/2)\cos(\omega_1^i t/2) + n_z^i\sin(\omega_0 t/2)\sin(\omega_1^i t/2),\\
b_i = & \sin(\omega_0 t/2)\cos(\omega_1^i t/2) - n_z^i\sin(\omega_0 t/2)\cos(\omega_1^i t/2).
\end{align}
An additional exponential decay $C_b=\exp{\left((-t/T_2^*)^2\right)}$, originating from the $^{13}\text{C}$ spin bath, is introduced. 
The stretching exponent is 2, because the correlation time of $^{13}\text{C}$ spin bath is considered to be longer than the experiment time. 
The experiment is conducted under a magnetic filed of $B\approx \SI{700}{G}$, hence the nitrogen spin in the NV center is strongly polarized to $\ket{m_{^{14}\text{N}}=+1}$ via the level anti-crossing (LAC) effect. 
Therefore, the microwave are played in the $\ket{m_{^{14}\text{N}}=+1}$ subspace, with a detuning frequency $\Delta\approx$ 2$\pi\times\SI{400}{\kilo Hz}$. 
Moreover, a slight population of $\ket{m_{^{14}\text{N}}=0}$ is taken into account, so the final experimental result of NV coherence $C$ is modified by the oscillating envelope:
\begin{align}
C_O = p_{\text{N}}\ee^{\ii \Delta t}+(1-p_{\text{N}})\ee^{\ii (\Delta-A_{^{14}\text{N}}) t+\phi},
\end{align}
where $A_{^{14}\text{N}}\approx -$ 2$\pi\times\SI{2.165}{MHz}$ is the longitudinal hyperfine parameter of $^{14}\text{N}$ spin, $p_{\text{N}}$ is the population of the $\ket{m_{^{14}\text{N}}=+1}$ state, and $\phi$ is an unknown phase. 
Using the known hyperfine parameters of the twelve $^{13}\text{C}$ spins, these expressions are utilized to fit the FID experimental data, as shown in Fig.~\hyperref[fig:FID]{\ref*{fig:FID}(a)}.

The fit yields an additional dephasing time of $T_2^*\approx \SI{12.8(4)}{\micro \second}$, which appears unexpectedly short compared to the magnetic field stability limit.
Moreover, fitting the data with only the first two $^{13}\text{C}$ spins gives a $T_2^*\approx \SI{10.68(23)}{\micro \second}$. 
This would counterintuitively suggest that the distant $^{13}\text{C}$ bath has a much stronger dephasing effect than the third to twelfth proximal $^{13}\text{C}$ spins.
A more plausible explanation is: one (or more) $^{13}\text{C}$ spin with a moderate $A_\parallel$ value fails to be detected by the DD sequence. 
This is possible if the $^{13}\text{C}$ spin has an extremely weak $A_\perp$ coupling, which would diminish its characteristic signal in the DD experiment.
Based on the density functional theory (DFT) calculation of the $^{13}\text{C}$ hyperfine parameters on different lattice sites, this kind of $^{13}\text{C}$ spins are scarce, so it is safe to hypothesize that only one spin is missing.

When fitting the data by adding the thirteenth $^{13}\text{C}$ spin but no additional $T_2^*$, the fitted value of $A_\parallel^{13} \approx 2\pi\times\SI{32.3(9)}{kHz}$. 
Therefore, it is attributed to one of the closest-matching sites identified via DFT calculation, which predicts an $A_\parallel \approx 2\pi\times\SI{31.13}{kHz}$ and $A_\perp \approx 2\pi\times\SI{6.4}{kHz}$.
As demonstrated in Fig.~\hyperref[fig:FID]{\ref*{fig:FID}(a)}, 
without the phenomenological exponential decay, these thirteen nuclear spins still nicely generate the FID experimental behavior.

\section*{Numerical simulation}

\subsection{$^{13}\text{C}$ configuration}

When simulating the dynamics of the spin network, only the subsystem $\mathscr{S}$ consisting of $N$ strongly-interacting $^{13}\text{C}$ nuclear spins is taken into account, along with the effect of the $\text{NV}$ center and the laser. 
Though twelve $^{13}\text{C}$ spins have been identified by the DD sequence, one spin by the FID sequence, and their hyperfine parameters are obtained, the values of $A_\parallel$ and $A_\perp$ alone cannot uniquely determine the spins' locations relative to the NV center. Instead, owing to the $C_{3v}$ symmetry of the system, a given pair of $(A_\parallel, A_\perp)$ values typically corresponds to three or six equally possible lattice sites. 
Consequently, the choice of the exact $^{13}\text{C}$ configuration is not unique. 
Without loss of generality, first the location choice of the first $^{13}\text{C}$ spin is fixed by selecting one representative from its equivalent positions. 
For the remaining twelve individual spins, some may locate on distant lattice sites where the influence on the spin dynamics is negligible.
Considering that the computational complexity limits the maximum number of spins in simulations, we aim to select several strongly-interacting spins and a corresponding configuration that produces simulation results most consistent with the experimental data, as shown in Fig.~\hyperref[fig:config]{\ref*{fig:config}} (the simulation methods are discussed in a latter subsection). 

\subsection{NV properties}

With the $^{13}\text{C}$ configuration determined, the only missing ingredient for simulation is the properties of the NV center.
Under green laser illumination, the NV is polarized to $\ket{m_S=0}$. 
The polarization degree is experimentally determined as $p_{m_S=0} \approx$ 0.9. 
For simplicity, the NV polarization dynamics is modeled within a master equation formalism, governed by four Lindblad operators:
\begin{align}
L_{0\rightarrow+1}&=\sqrt{\gamma_{0, +1}} \ket{m_S=+1}\bra{m_S=0},\ L_{+1\rightarrow 0}=\sqrt{\gamma_{+1, 0}} \ket{m_S=0}\bra{m_S=+1},\\
L_{0\rightarrow-1}&=\sqrt{\gamma_{0, -1}} \ket{m_S=-1}\bra{m_S=0},\ L_{-1\rightarrow 0}=\sqrt{\gamma_{-1, 0}} \ket{m_S=0}\bra{m_S=-1}.
\end{align}
When illustrated by laser, the phenomenological (de)polarization rates are extracted from experimental data,
\begin{align}
\gamma_{+1, 0}&=\gamma_{-1, 0}\approx \SI{10}{\micro \second}^{-1},\\
\gamma_{0, +1}&=\gamma_{0, -1}\approx \frac{1-p_{m_S=0}}{2} \gamma_{+1, 0}. 
\end{align}
When the laser is off, however, the NV dynamics is governed by the $T_1$ relaxation, thus,
\begin{align}
\gamma_{+1, 0}=\gamma_{-1, 0}=\gamma_{0, +1}=\gamma_{0, -1} = \frac{1}{3T_{1e}},
\end{align}
where $T_{1e} \approx \SI{5}{ms}$ is the experimentally observed NV longitudinal relaxation time. 
Moreover, simulations show that the NV fluctuation within the ground sublevels cannot fully explain the observed spin localization phenomena.
Phenomenologically, an extra dephasing operator is introduced for each nuclear spin $i$, induced by its interaction with the NV excited states ($^3E$):
\begin{align}
L_{i}^{\text{ext}}=\sqrt{\gamma_i} \sigma_z^i
\end{align}
with $\gamma_1=0.2\,\text{\textmu s}^{-1}$, and
\begin{align}
\gamma_i = \gamma_1 \times \left(\frac{A_z^i}{A_z^1}\right)^2,
\end{align}
where $A_z^i=\sqrt{(A_\parallel^i)^2+(A_\perp^i)^2}$. 
These dephasing operators are only effective when applying the laser pulses, which pump the NV to its excited states.

\subsection{Simulations of the autocorrelation function}

The simulations of $C_{zz}(t)$ are performed by numerically solving the Lindblad master equation, using the Lindblad operators introduced above, together with the Hamiltonian
\begin{align}
H=H_\text{n}+H_\text{hyp},
\end{align}
where $H_\text{n}$ is the nuclear Hamiltonian:
\begin{align}
H_\text{n}=\sum_{i=1}^N \omega_L I_z^i + \sum_{i< j}\frac{\mu_0}{4\pi r_{ij}^3}\left[\mathbf{I}_i \cdot \mathbf{I}_j -3(\mathbf{I}_i \cdot \mathbf{n}_{ij})(\mathbf{I}_j \cdot \mathbf{n}_{ij}) \right],
\end{align}
$r_{ij}$ is the distance between spin $i$ and spin $j$, and $\mathbf{n}_{ij}$ is the unit vector pointing from spin $i$ to spin $j$. 
Transforming $H_\text{n}$ to the interaction picture generates the $H_\text{n,int}$ in the main text, but the full Hamiltonian $H_\text{n}$ is preserved for simulating the AZE behavior.
To mitigate the stringent requirement for extremely small simulation time steps caused by the high-frequency term $\omega_L I_z^i$, the Strang splitting method is employed.

The initial state of $N$ nuclear spins is the tensor product of a pure state ($\ket{\uparrow}$ or $\ket{\downarrow}$) of the target spin and the maximally mixed state of the rest $(N-1)$ spins.
The NV charge state is always $\ket{\text{NV}^-}$; its initial spin state is the maximally mixed state, $\mathbb{I}/3$. 
The effect of $\ket{\text{NV}^0}$ is not taken into account for the following reasons.
When observing the unmonitored thermalization, an NV charge-state readout is performed during the initialization step and only the $\ket{\mathrm{NV}^-}$ events are post-selected. 
In the experiments with periodic measurements, although no charge-state readout is performed, the laser pulses rapidly drive the NV charge state transition between $\ket{\text{NV}^-}$ and $\ket{\text{NV}^0}$ on a timescale much faster than the nuclear $T_1$.
Therefore, the effect of $\ket{\text{NV}^0}$ is considered to be a multiplicative factor that only affects the long-time decay envelope.

The experiment is conducted under a high magnetic field, so the flip-flop terms among $N$ spins are the dominant Hamiltonian when the single-spin AZE is negligible. 
This leads to an undesired effect: the flip-flop process preserves the total nuclear spin number (e.g., 1), so the $C_{zz}$ value in some final equilibrium states is not $0$, but rather $\frac{1}{N}$. 
In real experiment, however, the environment $\mathscr{E}$ composed of the surrounding weakly-interacting nuclear spins also influences $\mathscr{S}$, thus the total nuclear spin number of $\mathscr{S}$ “leaks” to $\mathscr{E}$ and eventually reaches 0. 
In order to compensate for this mechanism and match simulation results more closely to experimental data, sometimes a transformation to the original simulation results is implemented:
\begin{align}
C_{zz; final} = C_{zz} - \frac{1}{N}((1-C_{zz})\times\frac{N}{N-1})^\beta
\end{align}
where the parameter $\beta \geq 1$ quantifies how “weak” $\mathscr{E}$ is coupled to $\mathscr{S}$. 
$\beta=1$ corresponds to when this coupling is approximately as strong as the internal coupling of $\mathscr{S}$, whereas $\beta \rightarrow +\infty$ indicates it being extremely weak, so that $C_{zz}$ already equilibrates at $\frac{1}{N}$ when the spin number “leaking” between $\mathscr{S}$ and $\mathscr{E}$ begins to happen.

Beyond the simulations already shown in the main text, we also ran simulations for other cases, each involving different effects.
The localization phenomena induced by global weak measurements is simulated in Fig.~\hyperref[fig:lasersimu]{\ref*{fig:lasersimu}}. 
Fig.~\hyperref[fig:AZE]{\ref*{fig:AZE}} demonstrates simulations of the additional effects, i.e., the anti-Zeno effect driven by $A_\perp$ coupling, and the $T_1$ shortening driven by the increased laser-induced polarization of the $\ket{m_S=0}$ state. 
Since the simulation of high-measurement-frequency region is computationally more complex, we only use 6 spins when simulating Fig.~\hyperref[fig:AZE]{\ref*{fig:AZE}(a)} and Fig.~\hyperref[fig:AZE]{\ref*{fig:AZE}(b)}, neglecting the less-effective No.\,7 and No.\,13 spin in the list of Fig.~\hyperref[fig:config]{\ref*{fig:config}}.

\subsection{Simulations of mutual information and entanglement entropy}

Since the mutual information $I_\text{A,B}$ obtained from the ensemble-averaged density matrix vanishes, the simulations must instead take a quantum-trajectory approach.
The simulations use the Hamiltonian $H_\text{n,int}$, and the effect of the NV electronic spin is modeled by the effective global weak measurement introduced in an earlier section. The effective Lindblad operators for each nuclear spin $i$ are
\begin{align}
L_i^\uparrow &= \sqrt{\gamma_\text{global}}\ket{\uparrow}_i \bra{\uparrow}_i,\\
L_i^\downarrow &= \sqrt{\gamma_\text{global}}\ket{\downarrow}_i \bra{\downarrow}_i,
\end{align}
where $\gamma_\text{global}=(1-p_{m_S=0})\lambda$, with $\lambda$ chosen as the laser pulse repetition frequency. Only $H_\text{n,int}$ and these Lindblad operators are included in the simulations.
The same nuclear spin configuration as in Fig.~\hyperref[fig:config]{\ref*{fig:config}} is adopted when the simulated system has 8 spins, and the simulation of 6 spins neglects the less-effective No.\,7 and No.\,13 spin. For the simulation of 10 spins, two extra spins are included: the eighth and ninth spin in the chart of Fig.~\hyperref[fig:DD]{\ref*{fig:DD}(d)}, with the lattice sites of 
$\vec{R_8}=(-7.5639,\, -2.9113,\, 3.5873)\,\text{\AA}$ and $\vec{R_9}=(2.5214,\, -7.2790,\, 3.5868)\,\text{\AA}$.
% $\vec{R_4}=(7.5667, \,5.8249, \,9.7651)\,\text{\AA}$ and $\vec{R_9}=(-7.5645, \,1.4559, \,3.5868)\,\text{\AA}$.

The subsystems A and B are each chosen to consist of a single nuclear spin.
Spin A always denotes the target spin.
Because the dipolar couplings are strongly inhomogeneous, the peak position $\gamma^*$ depends on the choice of A and B: a more strongly coupled pair remains correlated up to stronger measurements and therefore peaks at a larger $\gamma$. 
To give a conservative estimate of the criterion for our spin network to be in the localized regime, the spin B in Fig.\,3(b) is chosen to be the No.\,6 spin in the list of Fig.~\hyperref[fig:config]{\ref*{fig:config}}, which makes the chosen pair  one of the most strongly coupled pairs in our configuration. 
For Fig.~\hyperref[fig:MI]{\ref*{fig:MI}}, the spin B is the No.\,2 spin in Fig.~\hyperref[fig:config]{\ref*{fig:config}}, which is relatively weakly coupled to the remaining spins in the simulated network and demonstrates a lower critical $\gamma^*$ value.

All simulations take averages over 4000 quantum trajectories and over the time intervals $t \in[100, 200]\,\text{ms}$.
Each trajectory starts from the Néel state: 
\begin{align}
\ket{\psi(0)}=\ket{\uparrow \downarrow \uparrow \downarrow \cdots \uparrow \downarrow}.
\end{align}

However, whereas the qubit ordering is well defined in previous studies of random unitary circuits and one-dimensional fermion or boson chains, no such natural ordering exists in our system, which consists of a three-dimensional arrangement of spins with widely varying coupling strengths. We therefore place spins A and B in the first two positions of the ordering and randomize the order of the remaining spins independently in each trajectory.

The simulations for the entanglement entropy $S(t)$ (Fig.~\hyperref[fig:entropy]{\ref*{fig:entropy}}) follow the same procedure as that for the mutual information, except that the spin ordering is completely random in each trajectory and the subsystem always consists of the first half of all the eight spins. The time window over which $S(t)$ saturates is chosen to be $t\in[100, 200]\,\text{ms}$.

\section*{Comparison to a previous study}
This experiment is related to, yet distinct from, a previous study~\cite{Maurer2012}, where intense laser pulses were used to continuously fluctuate the NV spin.
In that regime, the laser illumination was so strong that it effectively diminished the many-body interactions among the $^{13}\text{C}$ spins; the target $^{13}\text{C}$ spin's dynamics was thus dominated solely by its direct coupling to the NV center.
The rapid NV spin fluctuation repeatedly interrupted the entanglement buildup between the NV electronic spin and the target spin, thereby prolonging the $^{13}\text{C}$ relaxation time via the single-spin QZE.
In contrast, the weak measurements in our experiment compete with the buildup of entanglement in the many-body nuclear spin system, giving rise to the observed measurement-induced spin localization. 
If we were to extend the parameter space of Fig.\,3(b) by implementing higher laser intensity, we anticipate that the dynamics would eventually overlap with the regime of ref~\cite{Maurer2012}, where the single-spin QZE dominates.

% \section*{Reference}
\bibliography{NV_abbr}

\newpage
\renewcommand{\thefigure}{S\arabic{figure}}

\begin{figure}
    \centering
    \begin{overpic}[width=1\linewidth]{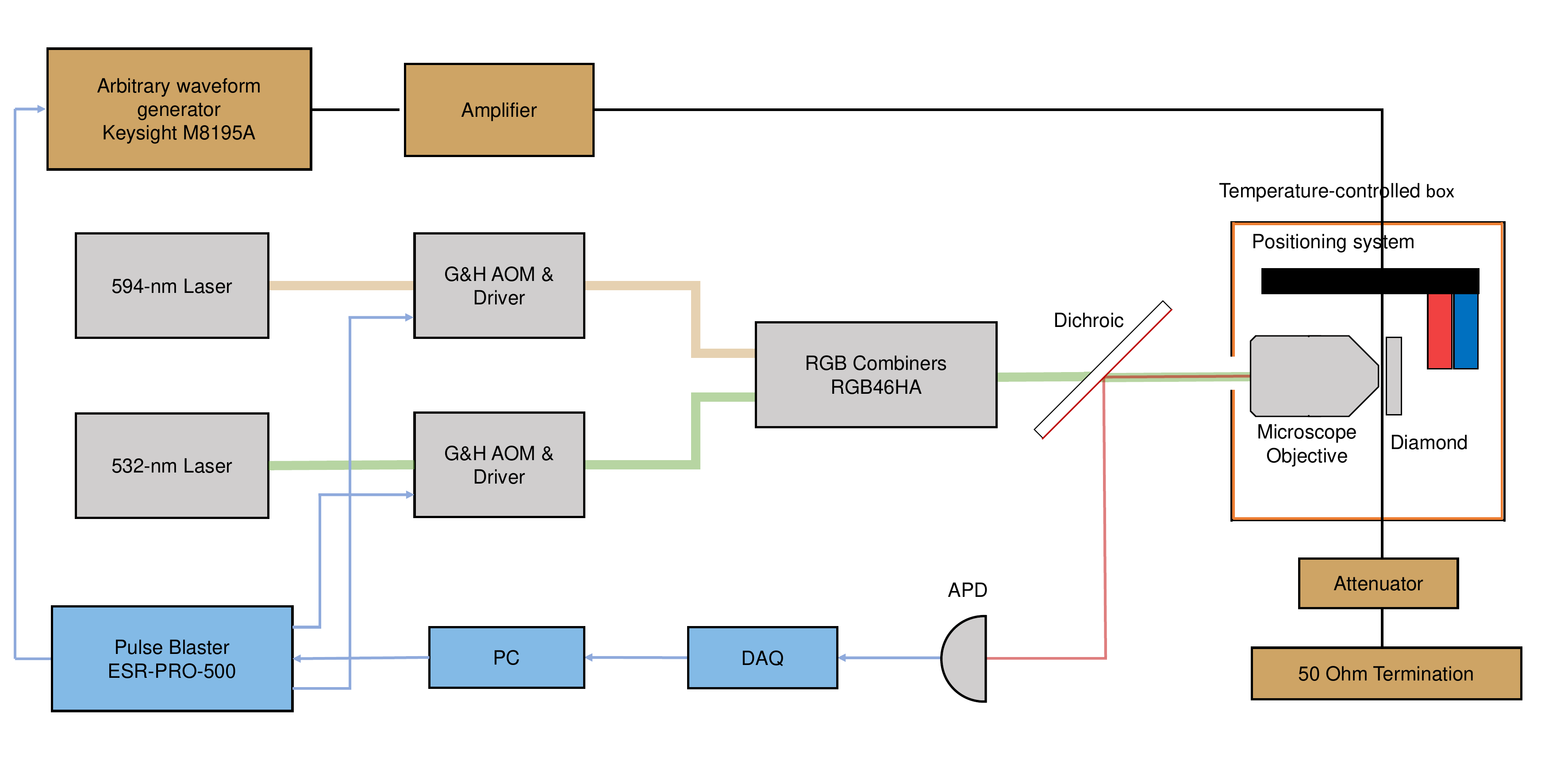}
    \end{overpic}
    \caption{
    Simplified experimental setup. 
    The setup mainly consists of microwave circuits (brown), optical systems (gray), synchronization systems (blue), and a diamond platform inside a temperature-controlled box.
    The blue, black and red lines carry TTL signals, MW pulses and phonon-sideband fluorescence, respectively.
    The thick yellow and green lines represent the corresponding optical beams.    
    }
    \label{fig:setup}
\end{figure}

\begin{figure}
    \centering
    \begin{overpic}[width=0.5\linewidth]{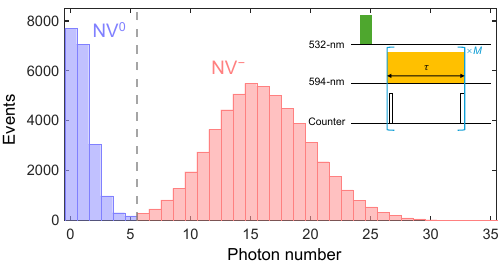}
    \end{overpic}
    \caption{
    Single-shot readout of NV charge state.
    Histogram of charge state readout based on the threshold method, showing 75000 readout events. 
    The subplot shows the sequence of charge state single-shot readout.
    Each event is the result of summing the photon counts over $M$ consecutive readout cycles. 
    During each cycle, the 594-nm laser illuminates continuously for a duration $\tau$, satisfying $M\tau=$ 4 ms.
    The gray dashed line denotes the threshold used to determine the NV charge state.
    }
    \label{fig:charge}
\end{figure}

\begin{figure}
    \centering
    \begin{overpic}[width=1\linewidth]{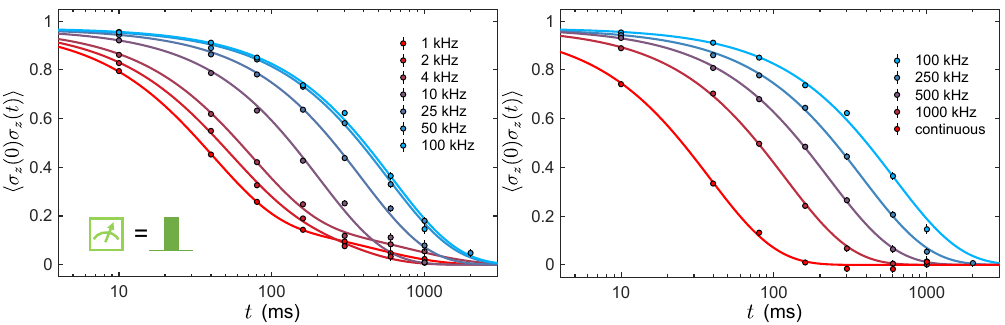}
    \put (0.5, 31.5) {\figfont{(a)}}
    \put (50.5, 31.5) {\figfont{(b)}}
    \end{overpic}
    \caption{
    Global measurements on the $^{13}\text{C}$ spin system.
    (a) Localization phenomena induced by global weak measurements. Each measurement only contains a 300-\text{ns} laser pulse.
    The spin system becomes more localized as the pulse repetition frequency increases. 
    (b) The AZE of frequent measurements.
    }
    \label{fig:laser}
\end{figure}

\begin{figure}
    \centering
    \begin{overpic}[width=1\linewidth]{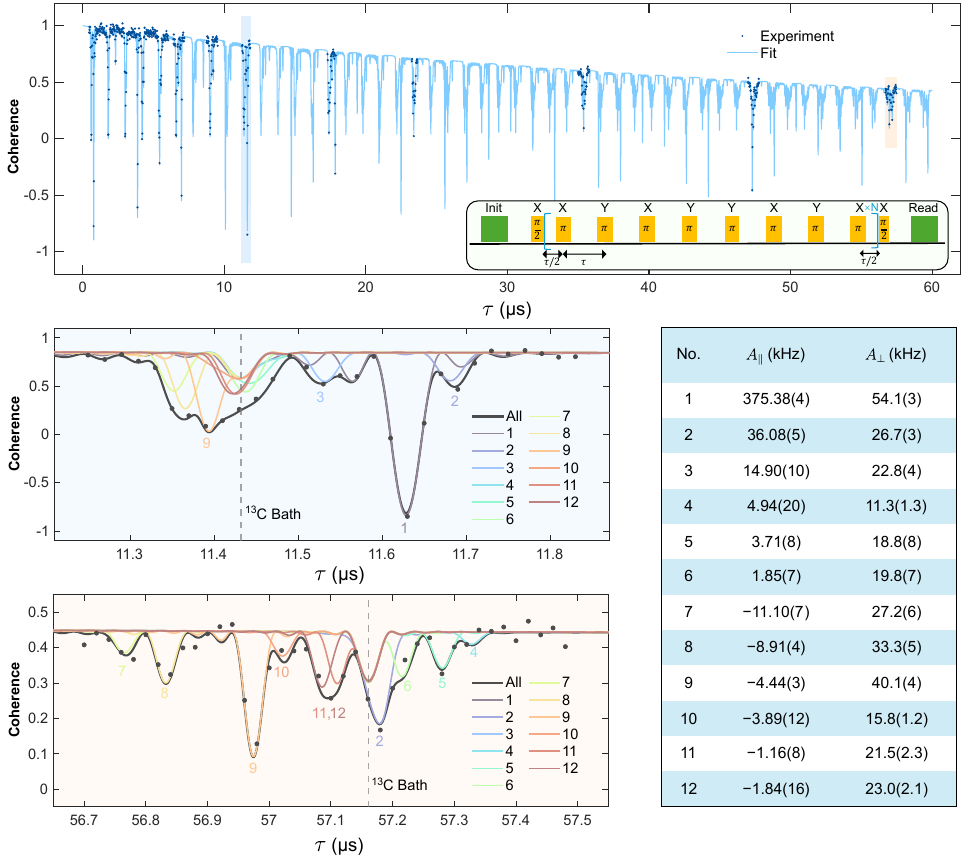}
    \put (0.5, 88) {\figfont{(a)}}
    \put (0.5, 54.5) {\figfont{(b)}}
    \put (0.5, 27.5) {\figfont{(c)}}
    \put (64, 54.5) {\figfont{(d)}}
    \end{overpic}
    \caption{
    Resolving $^{13}\text{C}$ nuclear spins by DD sequence.
    (a) The decoherence profile of the NV electron spin by applying the XY8-32 sequence.
    The results are fitted by combining the decoherence contributions from twelve nuclear spins and the spin bath. 
    Inset shows the XY8 sequence.
    (b), (c) Expanded view of selected regions from (a) to resolve individual $^{13}\text{C}$ spins.
    The colored curves mark the contribution of the twelve individual $^{13}\text{C}$ spins. 
    The vertical gray dashed lines correspond to the peak centers of the residual spin bath signal.
    (d) Fitting results of the $^{13}\text{C}$ hyperfine parameters.
    }
    \label{fig:DD}
\end{figure}

\begin{figure}
    \centering
    \begin{overpic}[width=1\linewidth]{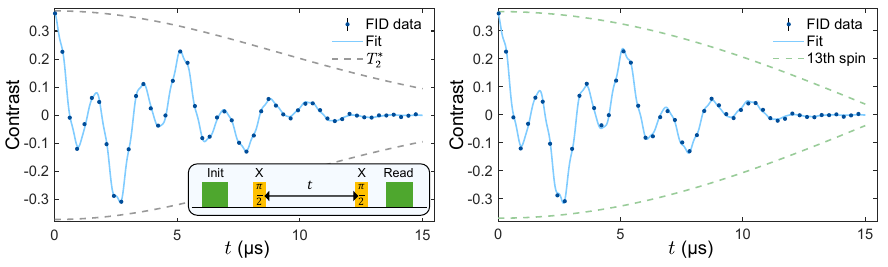}
    \put (1, 29) {\figfont{(a)}}
    \put (50, 29) {\figfont{(b)}}
    \end{overpic}
    \caption{
    An additional $^{13}\text{C}$ nuclear spin.
    (a) The decoherence profile of the NV electron spin by applying the Ramsey sequence.
    The results are fitted by combining the decoherence contributions from twelve $^{13}\text{C}$ spins and residual spin bath. 
    The decoherence contribution of spin bath is taken to be an additional exponential decay with a characteristic time of $T_2^*$, denoted as the dashed lines.
    (b) 
    The same experimental data is fitted by the decoherence contributions from twelve $^{13}\text{C}$ spins and one additional $^{13}\text{C}$ spin, whose hyperfine coupling parameters are $A_\parallel = \SI{31.13}{kHz}$ and $A_\perp = \SI{6.4}{kHz}$.
    The dashed line shows the decoherence contribution of the additional $^{13}\text{C}$ spin.
    }
    \label{fig:FID}
\end{figure}

\begin{figure}
    \centering
    \begin{overpic}[width=0.5\linewidth]{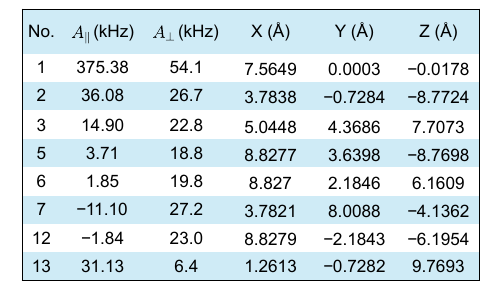}
    \end{overpic}
    \caption{
    Configuration of eight chosen $^{13}\text{C}$ spins and their hyperfine parameters.
    The coordinate axes are defined according to the following principle. 
    The origin is placed on the line connecting the nitrogen (N) and vacancy (V) of the NV center, on the side of V opposite to N, at a distance of $\SI{0.0171}{\text{Å}}$ from V.
    The Z-axis is aligned with the NV symmetry axis, with its positive direction pointing from V to N. 
    The Y-axis is chosen as one of the three equivalent directions such that one of the three carbon atoms nearest to V has an X-coordinate of zero. 
    Finally, the X-axis is determined to satisfy the right-hand rule with respect to the Y- and Z-axes.
    }
    \label{fig:config}
\end{figure}

\begin{figure}
    \centering
    \begin{overpic}[width=0.5\linewidth]{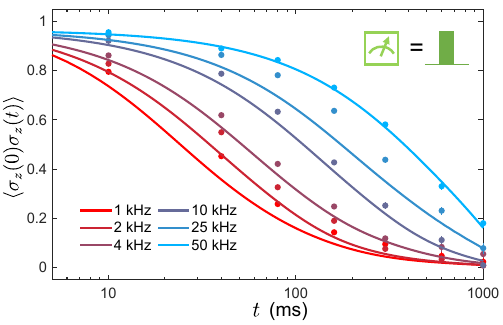}
    \end{overpic}
    \caption{
    Simulations of spin localization phenomena induced by global weak measurements.
    The data points are experimental results at the corresponding frequencies, shown for comparison.
    }
    \label{fig:lasersimu}
\end{figure}

\begin{figure}
    \centering
    \begin{overpic}[width=1\linewidth]{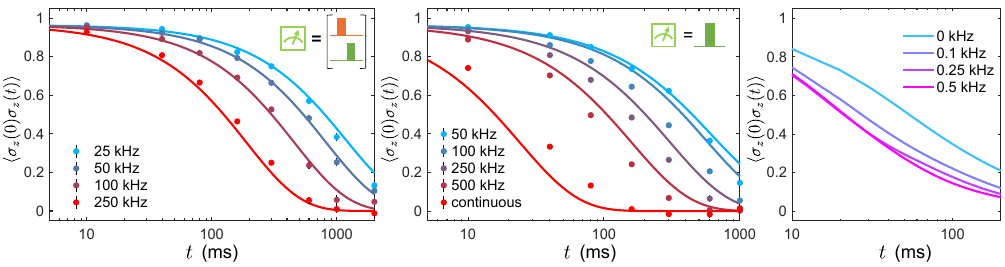}
    \put (0.5, 26) {\figfont{(a)}}
    \put (38, 26) {\figfont{(b)}}
    \put (74.5, 26) {\figfont{(c)}}
    \end{overpic}
    \caption{
    Simulations of additional effects.
    (a), (b) Simulation of the anti-Zeno effect induced by projective measurements and global weak measurements, respectively. 
    (c) 
    Simulation of the $T_1$ shortening in experiments with global weak measurements. 
    }
    \label{fig:AZE}
\end{figure}

\begin{figure}
    \centering
    \begin{overpic}[width=0.5\linewidth]{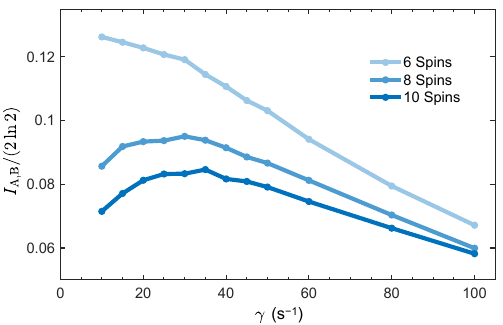}
    \end{overpic}
    \caption{
    Steady-state values of mutual information $I_\text{A,B}$ of two subsystems of the spin network at different measurement frequencies.
    }
    \label{fig:MI}
\end{figure}

\begin{figure}
    \centering
    \begin{overpic}[width=0.5\linewidth]{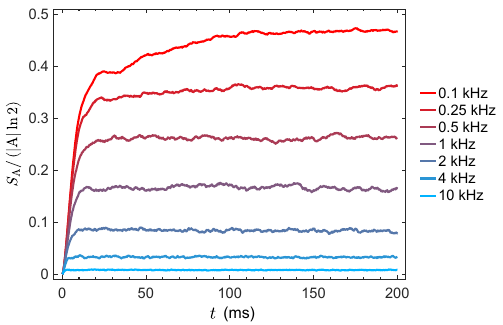}
    \end{overpic}
    \caption{
    Simulations of the entanglement entropy growth under different measurement frequencies. $N_\text{spin}=8$.
    }
    \label{fig:entropy}
\end{figure}